\documentclass{article}

\usepackage[preprint]{neurips_2026}

\usepackage[utf8]{inputenc} %
\usepackage[T1]{fontenc}    %
\usepackage{hyperref}       %
\usepackage{url}            %
\usepackage{booktabs}       %
\usepackage{amsfonts}       %
\usepackage{nicefrac}       %
\usepackage{microtype}      %
\usepackage{xcolor}         %
\usepackage{amsmath} 
\usepackage{enumitem}

\title{
Generative AI Advertising as a Problem of Trustworthy Commercial Intervention
}

\author{
  Jingyi Qiu \\
  School of Information \\
  University of Michigan \\
  \texttt{jaqiu@umich.edu} \\
  \And
  Qiaozhu Mei \\
  School of Information \\
  University of Michigan \\
  \texttt{qmei@umich.edu} \\
}

\begin{document}

\maketitle

\begin{abstract}
Major deployed generative AI advertising systems preserve a visible boundary between commercial content and AI-generated responses. Yet empirical research shows that ads woven directly into large language model (LLM) outputs often go undetected by users. We argue that generative AI fundamentally changes advertising: rather than placing products into discrete slots, it enables interventions on the generative process itself, which induce commercial influence through less observable channels. This reframes generative AI advertising as a problem of trustworthy intervention rather than content placement. We introduce a taxonomy organized by influence tier, corresponding to interventions on progressively more latent variables: product mentions, information framing, behavioral redirection, and long-term preference shaping; and show how these tiers instantiate across modalities and system architectures, including retrieval-augmented generation and agentic pipelines where upstream decisions can sharply constrain downstream outcomes. Both major deployed systems and designed mechanisms concentrate on the most observable and easiest-to-govern tier, while the forms of commercial influence most consequential for user autonomy remain poorly understood and lack frameworks for detection, measurement, or disclosure. The central challenge is whether commercial influence in generative systems can be made trustworthy, i.e., attributable, measurable, contestable, and aligned with user welfare. 
\end{abstract}

\section{Introduction}
Monetization is now a strategic priority for major AI systems. Google has made existing Search, Shopping, and Performance Max campaigns eligible for ads placed around AI Overviews~\citep{google2026aioverviews}; Microsoft is extending Performance Max and AI Max into multiple Copilot products %
~\citep{microsoft2024adsvoice,microsoft2026aimax}; and OpenAI is building commerce infrastructure around ChatGPT through its Agentic Commerce Protocol and Instant Checkout~\citep{openai2026commerce,openai2026productdiscovery}. LLM-mediated commerce is already measurable: a 12-month analysis of 973 e-commerce websites finds that organic ChatGPT referrals %
exhibit conversion rates and revenue per session above paid social~\citep{kaiser2026frontiers}. The question is no longer whether LLMs can be commercial interfaces, but how commercial influence will be integrated into them without undermining the trustworthiness of AI-generated outputs.

First-generation generative AI advertising is defined by restraint. Major documented deployments preserve a visible boundary between commercial content and AI-generated responses, using clearly labeled sponsored units visually separated from the organic answer~\citep{openai2026adshelp,openai2026testingads,microsoft2024adsvoice,google2026aioverviews}.\footnote{This paper characterizes publicly documented deployment patterns as of May 2026. } The mechanism design literature, by contrast, shows how quickly the design space moves beyond separated ads placement: token-level auctions, RAG-based segment auctions, sponsored LLM summaries, and aggregation of advertiser preferences over generated replies\citep{duetting2024mechanism,hajiaghayi2024ad,dubey2024auctions,soumalias2024truthful}. These proposals make explicit the allocative problem in LLM advertising, but they also reveal why production systems remain cautious: once interventions that induce commercial influence enter the generation process, the boundary between organic responses and ads becomes harder to localize, disclose, and audit.

Generative AI opens channels of commercial influence for which explicit attribution does not apply. Models can reshape emphasis, framing, and evidentiary context without displaying any sponsored object; generative engine optimization can increase the visibility of selected sources in model responses~\citep{aggarwal2024geo}; LLM-generated framings can shift attitudes even when factual content is held constant~\citep{tohidi2025rethinking}; and shopping agents exhibit position biases, reward platform endorsements, and allow seller-side description changes to shift market share~\citep{allouah2026your}. Commercial influence can be generated not only through what the model says, but through which sources it surfaces, which options it compares, and which tools or actions it selects. Empirical work shows these mechanisms are both feasible and behaviorally consequential: ads woven into chatbot responses often go undetected, can match or outperform ad-free outputs on perceived helpfulness, and admit adversarial amplification through prompt or content manipulation~\citep{tang2025ads,salvi2026commercial,meguellati2024good,meguellati2025llm,lin2025llm,nestaas2024adversarial}. %

\textbf{We argue that generative AI advertising should be studied as a problem of trustworthy intervention in the generative process, not as a problem of content placement.} This trust concern is especially salient in generative AI: unlike search results or social feeds, generative AI systems are used in open-ended, highly personalized, and potentially sensitive context %
and to support high-stake decisions. Framing the problem around trustworthiness shifts attention from \emph{where}  commercial signals appear to \emph{how} commercial interventions operate in the generative process, \emph{what} influence they induce, and \emph{how} it can be detected, measured, attributed, and where appropriate, disclosed or constrained. 

This does not imply that all commercial influence is harmful: under bounded attention and information overload, users may benefit from systems that surface relevant products, summarize credible evidence, and reduce information frictions in acting on their preferences. The central problem is distinguishing welfare-enhancing assistance from conflicted influence. 

We call this the problem of \emph{trustworthy commercial intervention}: whether commercial interventions in the generative process can be made normatively permissible, identifiable, measurable, and contestable. §~\ref{sec:taxonomy} formalizes this through a causal framework and shows why these requirements become harder as influence moves from product mentions to framing, behavioral redirection, and preference shaping. This framing also aligns platforms' commercial incentives with users' epistemic interests. Platforms that can characterize the forms of influence operating in their generative processes are better positioned to preserve the trust that sustains user engagement; those that cannot must choose between rejecting advertising entirely and accepting reputational risk from influence they cannot characterize or control.

This paper makes three contributions:
\begin{enumerate}[nosep]
\item A \textbf{taxonomy of generative AI advertising} organized by influence tier: product mentions (Tier~1), content framing (Tier~2), behavioral redirection (Tier~3), and long-term preference shaping (Tier~4). These tiers correspond to interventions on progressively more latent variables in the generative process, with modality and commercial-incentive structure as cross-cutting considerations. We also analyze how such interventions induce commercial influence in RAG and agentic pipelines (§\ref{sec:taxonomy}).

\item An \textbf{analysis of trustworthiness challenges at each tier}, organized around three operational requirements: identification, influence estimation, and contestability. We show where current detection methods fail, where disclosure becomes non-local or insufficient, and how cascade effects in agentic systems amplify upstream commercial interventions 
(§\ref{sec:taxonomy}).

\item A \textbf{mapping of open problems} across tiers, identifying technical and institutional capabilities required for trustworthy integration of commercial influence in generative systems.
\end{enumerate}

\section{The Current Landscape}
\label{sec:landscape}

\subsection{What are Deployed in Industry}

The first generation of advertising and commerce surfaces in AI assistants is now live, and official platform documentation shows a cautious convergent pattern: clearly labeled sponsored units, visually separated from the organic response, with placement boundaries and relevance matching rather than disclosed paid insertion into generated prose~\citep{openai2026adshelp,openai2026testingads,microsoft2024adsvoice,google2026aioverviews}. Table~\ref{app:industry} documents the official characterizations across major surfaces. 

Platforms differ in how they combine assistant access, advertising, and subscription monetization. For example, Google integrates ads into AI Overviews by default for all users, drawing from existing Search, Shopping, and Performance Max campaigns~\citep{google2026aioverviews}. OpenAI tests ads for eligible Free and Go users while keeping paid tiers ad-free~\citep{openai2026adshelp}. Anthropic takes a stronger position, stating that Claude will remain ad-free and that responses will not be influenced by advertisers or include unsolicited third-party product placements~\citep{anthropic2026spacetothink}.

\begin{table}[h!]
\centering
\caption{Officially documented advertising, commerce, and ad-free positions in major AI surfaces as of May 2026. The unit of comparison is the deployed user-facing surface, not the company or underlying model family.}
\label{app:industry}
\footnotesize
\setlength{\tabcolsep}{4pt}
\begin{tabular}{@{}p{1.6cm}p{1.8cm}p{7.2cm}p{2.4cm}@{}}
\toprule
\textbf{Surface} & \textbf{Documented commercial feature} & \textbf{Official characterization} & \textbf{Classification} \\
\midrule
ChatGPT &
Sponsored ad units &
Ads may appear below responses for eligible Free and Go users. OpenAI states that ads are clearly labeled, visually separated from ChatGPT's response, and do not influence answers. Its advertiser documentation describes CPM and CPC buying options. &
Paid ads; separated from response~\citep{openai2026adshelp,openai2026testingads,openai2026chatgptadsbasics} \\
\addlinespace

Microsoft\\ Copilot &
Copilot ad block &
Microsoft states that ads in Copilot appear below the organic response, are triggered using the whole conversation within a session, and are introduced by an ``ad voice'' explaining the connection to the conversation. Microsoft also describes AI Max as delivering ads on AI surfaces such as Copilot Search and Copilot Answers. &
Paid ads; separated from organic response~\citep{microsoft2024adsvoice,microsoft2026aimax} \\
\addlinespace

Google \\AI Overviews &
Search and Shopping ads &
Google states that existing Search, Shopping, Performance Max, and App campaigns are eligible to show ads above, below, and, in eligible markets, within AI Overviews; ads use existing auction ranking systems and signals. &
Paid ads in AI-search surface~\citep{google2026aioverviews} \\
\addlinespace

Meta AI &
Shopping mode and cross-app ad personalization &
Meta states that Meta AI can support shopping-oriented product discovery. Meta also states that interactions with AI at Meta may be used to personalize content and ad recommendations across Meta platforms. The official sources do not disclose paid in-chat placement or advertiser ranking within Meta AI responses. &
Product discovery and ad targeting; paid in-chat placement not disclosed~\citep{meta2026musespark,meta2025airecommendations} \\
\addlinespace

Perplexity &
Instant Buy and product cards &
Perplexity states that U.S. users can search for products, view product cards, and purchase directly through Perplexity. Its help page states that product listings are algorithmically ranked and that advertisers cannot pay to appear in related products. &
Commerce/product discovery; not sponsored~\citep{perplexity2026instantbuy} \\
\addlinespace

Claude &
No ads in Claude conversations &
Anthropic states that Claude will remain ad-free: users will not see sponsored links adjacent to Claude conversations, Claude responses will not be influenced by advertisers, and Claude will not include unsolicited third-party product placements. &
Explicitly ad-free~\citep{anthropic2026spacetothink} \\
\bottomrule
\end{tabular}
\vspace{0.35em}
\begin{minipage}{0.96\linewidth}
\footnotesize
\emph{Note.} We include only mainstream user-facing AI surfaces for which official public documentation describes a paid advertising surface, a commerce/product-discovery surface, or an explicit ad-free commitment. We therefore exclude surfaces for which we found no official public documentation of an in-assistant paid advertising product, including Gemini Apps, Grok, Mistral Le Chat, DeepSeek, Z.AI, and Meta Llama. 
\end{minipage}
\end{table}

\paragraph{How advertisers actually bid.}

To our knowledge, no publicly documented production system discloses an open auction for inserting sponsored content into generated response text. Instead, current systems adapt conventional impression-, click-, and campaign-level advertising infrastructure to AI surfaces. OpenAI's documentation describes ChatGPT Ads as supporting CPM and CPC buying options through a relevance-weighted, second-price auction, with selection based primarily on relevance to the context and intent of the conversation, using inputs such as context hints, landing page, ad title, and ad copy~\citep{openai2026chatgptadsbasics}.\footnote{CPM, or cost per mille, denotes the price per thousand ad impressions. CPC, or cost per click, denotes payment when a user clicks an ad.} Personalized ads may also draw on signals such as past chats and ad interactions~\citep{openai2026adshelp}. Google AI Overviews and Microsoft Copilot do not expose separate per-AI-surface bidding mechanisms; instead, eligible ads are drawn from existing campaign products such as Performance Max, with platform ML systems handling cross-surface optimization~\citep{google2026aioverviews,microsoft2024adsvoice,microsoft2026pmax,microsoft2026aimax}.\footnote{Performance Max is an automated campaign type in which advertisers specify goals, budgets, and creative assets while the platform optimizes bidding and delivery across eligible inventory.} 

The commercial frontier is also extending beyond conventional ad slots toward AI-mediated product discovery, checkout, and agentic commerce, including OpenAI's Agentic Commerce Protocol and Instant Checkout~\citep{openai2026commerce,openai2025instantcheckout}, Google's AI Mode shopping and Universal Commerce Protocol ~\citep{google2025aimodeshopping,google2026ucp}, Microsoft's Copilot Checkout and Brand Agents~\citep{microsoft2026copilotcheckout}, Perplexity's Instant Buy~\citep{perplexity2026instantbuy}, and Meta's shopping mode~\citep{meta2026musespark}.

\subsection{What are Proposed in Literature}

In contrast to industry's caution, a growing body of scholarly work explores mechanisms for incorporating advertising into AI-generated outputs. These include token-level auctions over generated content~\citep{duetting2024mechanism}, segment-level RAG auctions in which ads are retrieved for discourse segments according to bid and relevance~\citep{hajiaghayi2024ad}, post-generation genre-based VCG %
ad insertion~\citep{xu2026ad}, learning-based generative auctions that jointly optimize allocation and generation~\citep{zhao2025llm}, placement auctions for LLM summaries~\citep{dubey2024auctions}, truthful aggregation of advertiser preferences over LLM-generated replies~\citep{soumalias2024truthful}, and broader conceptual or governance frameworks~\citep{feizi2023online,wu2025advertising}. \citet{hu2025gem} introduce a benchmark for ad-injected response generation, comparing prompt-based ad injection with approaches that first generate an ad-free response and then insert or refine ad content. Detection methods remain similarly concentrated on the explicit setting: fine-tuned sentence transformers can identify generated native ads with high precision and recall~\citep{schmidt2024detecting}, but this evidence concerns inserted ad text rather than subtler forms of framing, redirection, or preference shaping. 

Most of these proposals assume the advertiser wants a specific product mentioned, what we call \emph{Tier~1 advertising} and develop in §\ref{sec:taxonomy}. Some mechanisms operate at a distributional level, including aggregating advertiser-conditioned token distributions~\citep{duetting2024mechanism}, advertiser preferences over candidate replies~\citep{soumalias2024truthful}, or distributions over outputs~\citep{zhao2025llm}, and are better characterized as content framing (Tier~2) than as product placement. 

Adjacent literatures offer useful tools but do not yet provide a framework for generative AI advertising. Native-advertising research studies disclosure recognition~\citep{wojdynski2016going}, dark-pattern work studies manipulative interface design~\citep{mathur2019dark}, recommender system research studies exposure allocation~\citep{singh2018fairness}, and search-neutrality work studies platform self-preferencing~\citep{de2014integration,motta2023self}. 
This literature informs the study of disclosure, manipulation, allocation, and platform incentives, but it does not directly address generative systems in which commercial influence %
is woven into %
the generative process.

\subsection{The Gap}

The forms of commercial influence most consequential for user autonomy do not operate through product mentions: framing, behavioral redirection, and preference shaping offer no natural correction mechanism since users cannot discount what they cannot perceive, and effects compound across interactions with no natural expiration. Yet both deployment and academic research concentrate on product placements, and existing evidence of deeper forms, including adversarial prompt steering~\citep{lin2025llm}, cognitive-bias manipulation~\citep{filandrianos2025bias}, and agent shopping biases~\citep{allouah2026your}, remains fragmentary feasibility demonstrations. What is missing is a systematic framework that organizes the full space by tier, modality, and mechanism while also identifying the corresponding trustworthiness challenges: what can be detected, attributed, disclosed, contested, or aligned with user welfare. This paper provides that framework.

\section{A Taxonomy of Generative AI Advertising}
\label{sec:taxonomy}

We propose a taxonomy of generative AI advertising organized by influence tier: interventions on progressively more latent variables in the generative process, with modality and bidding mechanism as cross-cutting considerations. As interventions move from visible outputs to latent framing, actions, and preferences, the trustworthiness requirements become harder to satisfy.

We define \emph{commercial influence} as the causal effect induced by interventions on the generative process: shifts in output distributions that serve an identifiable commercial interest beyond what would be produced under the model's baseline (no-intervention) distribution. Such interventions can arise from \emph{paid advertising} (external actors paying the platform to advance commercial interests), \emph{platform self-preferencing} (the platform favoring its own products without external payment), or \emph{third-party adversarial steering} (external parties exploiting prompts, retrieval documents, or product descriptions without platform cooperation). These produce similar observable effects but call for distinct governance responses: disclosure, antitrust enforcement, and platform security, respectively; our taxonomy applies to all three.

Classical computational advertising can be abstracted as finding the best match between a user $u$ in a context $c$ and a promoted object $x$, typically through allocation over discrete ad opportunities such as sponsored-search positions or display impressions~\citep{broder2008computational,edelman2007internet,yuan2013real}. In generative AI, $x$ need not appear as a discrete sponsored unit; it may be promoted indirectly through interventions in the generative process, shaping content, actions, and user states.

We introduce a unifying probabilistic framework with a causal interpretation. Let $x$ denote a promoted commercial object, $u$ the user preferences, $c$ the action context (e.g., conversation, query, or situation), and $X$ a cluster or category of products with $x \in X$. The core quantity of interest is the probability that product $x$ is recommended or selected, $P(x \mid u, c)$. We interpret advertising as an intervention $\pi$ on the generative process, yielding the interventional distribution $P(x \mid do(\pi),u,c)$. Commercial influence is defined as the resulting treatment effect relative to a no-intervention baseline:
\begin{equation} \mathrm{ATE}(\pi; u, c) = P(x \mid do(\pi), u, c) - P(x \mid do(\emptyset), u, c).
\end{equation}

\paragraph{Trustworthy commercial intervention.} This causal framing clarifies what it would mean for a commercial intervention to be trustworthy. Beyond a normative requirement that the intervention be ethically permissible, three operational requirements recur across tiers: \emph{identification} (users, auditors, or regulators can determine that an intervention occurred and which part of the generative process it affected); \emph{influence estimation} (the system supports counterfactual evaluation of how the response, action, or user state would have differed absent the intervention); and \emph{contestability} (users or auditors can discount, opt out of, reinterpret, constrain, or reverse the intervention's effects). The taxonomy below shows how these requirements become harder to satisfy as interventions move from observable outputs to latent framing, downstream actions, and long-term preferences. Each tier decomposes the outcome probability through a different intermediate variable, identifying a different pathway through which commercial intervention can change the probability of observing $x$.

\subsection{Influence Tiers}
\label{sec:tiers}

Table~\ref{tab:tiers} summarizes the four tiers: each row specifies what is promoted, the underlying intervention target, how such influence can be detected, and the corresponding ethical concerns, illustrating a progression from observable placement to latent preference shaping. %

\begin{table}[t]
\centering

\caption{
Four tiers of advertising influence in generative AI. Each tier corresponds to interventions on progressively more latent variables in the generative process. %
}
\small
\begin{tabular}{p{1.9cm} p{2.6cm} p{2.6cm} p{1.9cm} p{3cm}}
\toprule
\textbf{Tier} & \textbf{What’s Promoted} & \textbf{Intervention Target} & \textbf{Detection} & \textbf{Key Ethical Concerns} \\
\midrule
\textbf{1: Product Advertising} 
& A specific product or brand 
& Output selection $x$ 
& Surface-level identification (e.g., product matching) 
& Deception in a trusted interface; disclosure salience matters
\\

\textbf{2: Content Framing} 
& A product category or narrative framing
& Content distribution $P(X \mid u, c)$ 
& Counterfactual baseline comparison 
& Unobservable framing effects; omission cannot be easily detected or contested \\

\textbf{3: Behavioral Redirection} 
& A user action or downstream pathway
& Context transition $c'$ 
& Action attribution analysis 
& Reduced autonomy via behavioral nudging and action steering; difficult to distinguish
helpful choice architecture from promotion\\
\textbf{4: Preference Shaping} 
& A preference or mental model 
& Preference dynamics $u'$ 
& Longitudinal outcome audit 
& Preference shaping without localized intervention; difficult to attribute or govern \\
\bottomrule
\label{tab:tiers}
\end{tabular}

\end{table}

\textbf{Tier~1: Product Advertising.} The intervention operates directly on output selection, shifting $P(x \mid u, c)$ by determining which specific product is mentioned. A concrete product or brand appears explicitly in the response, either as a labeled placement or as an integrated recommendation. 

\emph{Across modalities.} Tier~1 manifests as a discrete commercial mention, but its surface form varies by interface. In text, a named product or brand appears within the generated response or labeled product cards adjacent to the generated answer. In image or video generation, it appears as a recognizable branded object, logo, or product within the generated scene. In code, it appears as a specific SDK, API, or library named in the generated output, with direct infrastructure and cost implications.

\emph{Bidding and mechanism design.} 
The auction logic for sponsored search~\citep{edelman2007internet} maps most cleanly onto Tier~1 placement: a discrete sponsored unit, typically ranked by bids and relevance and billed through measurable outcomes such as clicks or impressions. The literature on generative AI advertising remains concentrated near Tier~1, i.e., output-level commercial content. Rather than simply inheriting sponsored-search mechanisms such as generalized second-price (GSP) auctions~\citep{edelman2007internet,varian2007position}
or Vickrey--Clarke--Groves (VCG) mechanisms~\citep{vickrey1961counterspeculation,clarke1971multipart,groves1973incentives}, recent work adapts auction design to LLM-specific allocation objects, including tokens~\citep{duetting2024mechanism}, retrieved RAG segments~\citep{hajiaghayi2024ad}, summaries~\citep{dubey2024auctions}, post-generation genre-based insertions~\citep{xu2026ad}, distributions over generated outputs~\citep{zhao2025llm}, and advertiser preferences over candidate replies~\citep{soumalias2024truthful}. These proposals retain classical concerns with allocation, pricing, welfare, and incentive compatibility, but apply them to generative outputs where content and placement are not fixed in advance. 
Publicly documented deployments appear more conservative, adapting existing cost-per-click (CPC), cost-per-impression (CPM), and campaign-level advertising infrastructure to AI surfaces rather than disclosing new auctions for paid insertion into generated responses.
Despite these variations, all approaches share the assumption that the objective is to select a specific product to be mentioned, and billing is tied to impressions or clicks on a discrete sponsored unit.

\emph{Ethical concerns.} Even at Tier~1, where influence is most observable, user awareness remains limited. \citet{tang2025ads} find that 49\% of users cannot detect woven Tier~1 ads, while \citet{salvi2026commercial} report that LLM-driven persuasion nearly triples the rate of sponsored product selection. The core ethical issue is deception: integrating product promotion into a trusted information source can exploit users' expectation that AI assistants provide independent advice~\citep{sundar2008main,ftc2015nativeadvertising,logg2019algorithm,tang2025ads}.

\emph{Open problems.} Two challenges are central even at Tier~1. First, when a product mention is woven into the generated response, even a user who recognizes it cannot easily discount its influence because they do not observe the counterfactual: what product, comparison set, evidence, or recommendation would have appeared in the same position under the no-ad baseline. The central evaluation problem is therefore not only disclosure, but counterfactual response-quality auditing, which requires comparing the sponsored response against a no-intervention baseline rather than merely detecting whether a brand name appears. Second, the stakes of insertion quality vary sharply by domain: in medical, legal, or financial contexts, even an explicit product mention can distort decision-making if it changes which alternatives are considered, which evidence is emphasized, or which risks are omitted.

\textbf{Tier~2: Content Framing.} This corresponds to introducing a latent content variable $X$, such as a product \emph{category} or a narrative. Through the decomposition
$$P(x \mid u, c) = \sum_{X} P(x \mid X, u, c)\, P(X \mid u, c), $$
interventions on $P(X \mid u,c)$ shift the outcome indirectly by altering which categories or narratives are made salient. Rather than promoting a specific product, the system increases its likelihood by framing the response in ways that favor categories in which that product is prominent. Examples include \textit{category steering} (``you should invest in a good treadmill,'' paid for by a dominant treadmill manufacturer), \textit{information framing} (emphasizing the advantages of electric vehicles in a car-buying query, sponsored by the EV industry), and \textit{problem amplification} (suggesting vitamin deficiencies rather than lifestyle changes in a fatigue query, sponsored by supplement manufacturers).

A growing body of evidence demonstrates that Tier~2 influence is technically achievable and can affect both model outputs and user attitudes. On the supply side, Generative Engine Optimization~\citep{aggarwal2024geo} shows that content creators can increase source visibility in LLM-generated responses by up to roughly 40\% through strategies including the addition of citations, quotations, and quantitative statistics. Adversarial variants are more direct: \citet{nestaas2024adversarial} show that manipulated web content can make a target product 2.5$\times$ more likely to be recommended in a Bing camera-recommendation setting; \citet{lin2025llm} show that subtle synonym replacements in prompts can increase target-concept mentions by up to a 78\% difference; and \citet{filandrianos2025bias} show that cognitive-bias strategies embedded in product descriptions can alter LLM recommendation rates and rankings. On the user side, \citet{tohidi2025rethinking} demonstrate in a preregistered experiment that LLM-generated articles varying selective emphasis while holding factual accuracy constant significantly shift policy attitudes and emotional responses.

\emph{Across modalities.} Tier~2 manifests as shifts in content framing, such as selective emphasis or omission in text, biased result framing in search, aesthetic and compositional choices in image generation, and systematic recommendation of a class of tools over alternatives in code.

\emph{Bidding and mechanism design.} 
No canonical bidding mechanism currently prices Tier~2 framing directly, although recent LLM-ad auction proposals begin to approach this setting by allocating retrieved segments or optimizing distributions over generated outputs~\citep{hajiaghayi2024ad,zhao2025llm}. Standard CPM/CPC units do not naturally apply, as there is no discrete, attributable unit of influence to price. Instead, interventions operate on the content distribution $P(X \mid u, c)$, suggesting billing models based on expected lift in category salience, e.g., topic-based retainers or per-query contracts. Natural buyers are industry associations and market-dominant companies that benefit from category-level demand without needing to be named (e.g., ``Got Milk?''-style campaigns). The absence of clear attribution and immediate feedback challenges core assumptions of mechanism design, including measurable outcomes, independent valuation, and verifiable allocation.

\emph{Ethical concerns.} Tier~2 introduces a distinct risk: the influence is invisible by construction. There is no explicit product mention to flag and no localized span to label as ``Sponsored.'' Users cannot readily detect or contest the influence. A user asking about retirement planning may receive a response that subtly favors active investment management over index funds without recognizing that the framing itself might be biased. Such influence operates through selective emphasis and information omission, shaping decisions without providing a clear signal of commercial intent.

\emph{Open problems.} First, defining ``neutral baselines'' that are robust to model stochasticity and prompt sensitivity is essential for detecting shifts in $P(X \mid u,c)$, but distinguishing commercial framing from organic model biases is inherently ambiguous, as both operate through similar mechanisms of emphasis and omission. Second, existing evaluation metrics are largely omission-blind: they capture factual distortion but not the absence of relevant alternatives, highlighting the need for omission-aware quality metrics. Similar Tier~2 mechanisms are already being studied in non-commercial contexts, including LLM-driven persuasion in political and policy contexts~\citep{costello2024durably,bai2025llm}, and AI-augmented state-backed disinformation campaigns~\citep{wack2025generative}.
Bridging these domains may provide methodological tools and empirical benchmarks.

\textbf{Tier~3: Behavioral Redirection.} This corresponds to introducing a latent variable $c'$, representing a downstream user action, tool-mediated pathway, or new decision context. Through the decomposition
$$P(x \mid u, c) = \sum_{c'} P(x \mid u, c')\, P(c' \mid u, c), $$
interventions on $P(c' \mid u,c)$ shift the probability that $x$ is later recommended or selected by steering the user toward particular downstream actions or pathways. Rather than altering content directly, the system influences which downstream contexts are realized. Examples include suggesting specific links, follow-up queries, or tool invocations that guide users toward commercially relevant outcomes.

Behavioral nudging provides a natural mechanism for Tier~3 influence. Building on the choice-architecture view of nudges~\citep{thaler2009nudge,weinmann2016digital}, Tier~3 interventions need not change what information is true or false. Instead, they change the downstream action environment: which next steps are made salient, which actions are easiest to take, which options are presented as defaults, and which pathways require additional friction. This distinguishes Tier~3 from Tier~2 framing: Tier~2 changes the informational or narrative context, whereas Tier~3 changes the action pathways through which the user proceeds. In agentic systems, nudges may be implemented through tool choice, default filters, ranking parameters, or suggested actions.

Early empirical evidence suggests Tier~3 is already relevant in agentic commerce. \citet{allouah2026your} show that AI shopping agents penalize explicitly labeled ``Sponsored'' items while rewarding platform-level endorsements, suggesting that agentic intermediaries may discount conventional ad labels while amplifying signals that appear organic or authoritative. Related work finds that agents respond less to human-facing visual or emotional ad cues and more to structured, machine-readable signals~\citep{stockl2025ai,nitu2025machine}. Long-standing debates over search neutrality and platform self-preferencing~\citep{de2014integration,motta2023self} provide institutional precedent, while emerging models of agentic purchasing formalize how agents can shape consumer choice through preference elicitation~\citep{cao2026solicit}.

\emph{Across modalities.} Tier~3 appears as explicit suggestions of next steps in text and search (links, follow-up queries, actions), and takes a more powerful and less visible form in agentic settings: back-end tool selection (e.g., Expedia vs.\ Booking.com) that users do not observe, thereby constraining the entire downstream consideration set (§\ref{sec:pipelines}). Perplexity's reportedly phased-out ``Sponsored Follow-Up Questions'' provide one of the few deployed examples to date.

\emph{Bidding and mechanism design.} 
To the best of our knowledge, no formal mechanism design framework currently treats Tier~3 action or pathway selection as its primary allocation object. 
Existing LLM-ad mechanisms focus mainly on output-side allocation. For Tier~3, the natural unit of allocation is the action itself, i.e., the downstream behavioral outcome induced by the system. 
Existing pricing models such as cost-per-action (CPA) or revenue sharing provide the closest analogues, with payment triggered when the user follows the suggested action. The central challenge is attribution. Determining whether an action occurred because of the system’s intervention requires 
estimating a counterfactual path of user
behavior, substantially exceeding the difficulty of traditional attribution problems. 
Tier~3 therefore inherits the well-known difficulties of incrementality testing and multi-touch attribution, while adding a further complication: the relevant intervention may be embedded in conversational advice, tool selection, or choice architecture rather than in a discrete, observable ad impression.

\emph{Ethical concerns.} Tier~3 raises a distinct concern for user autonomy: influence operates not by distorting information, but by channeling behavior. A response can be factually accurate, balanced, and unbiased, yet still steer the user toward commercially favored actions. This dual nature, simultaneously helpful and commercially motivated, makes such influence difficult to recognize and contest, blurring the boundary between neutral assistance and promotion.

\emph{Open problems.} First, because behavioral redirection operates through downstream context transitions rather than visible response content, detection requires estimating whether a later action, tool invocation, or purchase was caused by an upstream suggestion rather than the user's baseline intent. This makes counterfactual attribution difficult, especially under privacy constraints. Second, disclosure is non-local: the same suggestion can be welfare-enhancing assistance or commercially motivated steering, making it hard to decide whether advice such as ``check reviews before buying'' is neutral guidance or promotion of a particular review platform. Agentic settings introduce further concerns around competition and self-preferencing, requiring safeguards for tool selection (§\ref{sec:pipelines}).

\textbf{Tier~4: Preference Shaping.} This corresponds to introducing a latent variable $u'$, representing an updated user preference state. Through the decomposition
$$P(x \mid u, c) = \sum_{u'} P(x \mid u', c)\, P(u' \mid u, c), $$
interventions on $P(u' \mid u,c)$ shift the outcome indirectly by influencing how user preferences evolve over time. Rather than affecting a single response, the system shapes future responses by reinforcing certain associations, priorities, or habits across interactions, e.g., repeatedly favoring particular brands, normalizing certain tradeoffs, or subtly shaping long-term beliefs about quality and value.

Direct evidence on deployed, commercially motivated Tier~4 preference shaping remains limited, but adjacent literatures show that LLM-mediated interaction can shift beliefs and attitudes in ways relevant to long-run preference formation: persuasive messages can shift policy attitudes with effects comparable to lay humans~\citep{bai2025llm}, interactive conversations can durably reduce conspiracy beliefs~\citep{costello2024durably}, and biased LLMs and writing assistants shift users' opinions even when the bias conflicts with prior partisanship~\citep{fisher2025biased,williams2026biased,sharma2024generative,matz2024potential}. Reverse-inference work shows why such influence can be targeted: \citet{chen2026ads} demonstrate that LLMs can infer private attributes such as political affiliation, employment status, and education level from passive ad exposure patterns at substantially lower cost than human profiling. These findings make Tier~4 empirically plausible and identify the quantities a trustworthy intervention regime must measure.

\emph{Across modalities.} Tier~4 operates through cross-session, cross-modal preference dynamics, including patterns across repeated conversations, persistent aesthetic biases in image generation, persistent tool or framework recommendations in code, and brand associations stored in agentic memory (§\ref{sec:pipelines}).

\emph{Bidding and mechanism design.} 
Standard per-impression billing is not well-defined for Tier~4, where influence unfolds over time rather than within discrete interactions. The natural analogue is long-term brand equity contracts, measured through instruments such as lift surveys, approaches long used in television advertising. To date, no academic mechanism design framework addresses this setting, reflecting the absence of clear units of allocation, attribution, or feedback.
A natural billing model could be long-term brand equity contracts measured by lift surveys, the same instruments used in television advertising for decades.

\emph{Ethical concerns.} Tier~4 raises the most fundamental concerns for user autonomy because the intervention targets preference formation rather than a discrete response, frame, or action. Unlike a product mention, preference shaping may leave no localized artifact for users to notice or contest. The ethical concern is not only opacity but cumulative dependence: repeated interaction may normalize particular tradeoffs, brands, sources, or decision criteria while making the resulting preference change appear endogenous to the user. This makes Tier~4 the setting in which identification, influence estimation, and contestability are most difficult, and where opt-out or preference-reset mechanisms become especially important.

\emph{Open problems.} First, detection requires longitudinal analysis: influence accumulates across sessions and cannot be identified from any single interaction, so designing study protocols that isolate the causal contribution of AI systems from external exposure is essential but difficult. Second, measuring preference drift in a privacy-preserving manner poses a fundamental tension between observability and user protection, and it is unclear whether existing advertising regulations (e.g., FTC guidelines, the EU's Digital Services Act) extend to diffuse, temporally extended influence.

\paragraph{Cross-tier synthesis.} Detection difficulty increases systematically across tiers as interventions move from observable variables ($x$ in Tier~1) to progressively more latent ones ($X$, $c'$, $u'$ in Tiers~2--4), though this hierarchy interacts with modality: detecting a Tier~1 branded object in a generated image may be more challenging than identifying a Tier~2 framing shift in text. The three operational requirements become correspondingly harder to satisfy at higher tiers, with Tier~4 making both identification and contestability longitudinal problems. Tiers can also stack within a single interaction: a response may simultaneously mention a product (Tier~1), frame its category favorably (Tier~2), and suggest a downstream action (Tier~3), while only the Tier~1 component remains easily detectable. 

\paragraph{Bidding and mechanism design across tiers.}
The auction logic of sponsored search maps most cleanly onto Tier~1: a discrete sponsored unit is ranked by bids and relevance and billed through measurable outcomes such as impressions, clicks, or conversions. Existing LLM-ad auction proposals largely remain close to this setting, even when the allocation object becomes LLM-specific, such as tokens, retrieved RAG segments, summary placements, post-generation insertions, distributions over generated outputs, or advertiser preferences over candidate replies. %

Higher tiers are less naturally captured by standard ad units. Tier~2 interventions operate on the content distribution $P(X \mid u,c)$, so pricing would have to approximate expected lift in category salience rather than price a discrete impression. Tier~3 interventions allocate downstream actions, making CPA or revenue-share models the closest analogues, but attribution becomes difficult because the relevant intervention may be embedded in conversational advice rather than a visible ad unit. Tier~4 interventions unfold over time, making long-term brand-equity measures such as lift surveys closer analogues than click-based billing. Thus, as influence moves from product placement to framing, action steering, and preference dynamics, the mechanism-design problem loses the clean assumptions of observable allocation, immediate feedback, and verifiable outcomes.

\subsection{Case Studies: Advertising in RAG and Agentic Pipelines}
\label{sec:pipelines}

The tiered framework characterizes where interventions act in the generative process; in practice, they are realized through concrete system architectures. Two dominant paradigms, retrieval-augmented generation (RAG) and agentic workflows, instantiate distinct pathways for introducing, amplifying, and propagating commercial influence. 

\paragraph{RAG pipeline.} A standard retrieval-augmented generation system operates in stages: embed query $q$, retrieve top documents $D$, optionally rerank, assemble prompt, and generate~\citep{lewis2020retrieval,gao2023retrieval}. Advertising enters at the retrieval stage similarly as sponsored search: ads stored in the same vector index as document relevance scores augmented by an advertiser-provided signal $\mathbf{s}(\cdot)$, biasing which documents are selected. In our notation (substituting initial context $c$ with query $q$):
\begin{equation}
P(x \mid u, q, \mathbf{s}) = \sum_{D} P(x \mid u, D, q) \cdot P(D \mid q, \mathbf{s}),
\end{equation}
where $D$ is the set of documents retrieved and $\mathbf{s}$ the bid-dependent scoring function. Interventions on $P(D \mid q, \mathbf{s})$ shift which information enters the model's context window. This primarily instantiates a Tier~2 intervention (content framing), but because the retrieved documents define the effective context $c' = (D, q)$, it can also induce Tier~3 effects by constraining downstream actions and recommendations. Unlike sponsored search, the intervention is invisible to the user: the response reads as organic even when retrieval was commercially biased. Detection would require auditing retrieval indices or ranking scores, which are typically unavailable to external auditors.

\paragraph{Agentic pipeline.} Agentic systems introduce a qualitatively different and more challenging surface for commercial intervention. We illustrate this using OpenClaw~\citep{openclaw2026,shan2026don}, an open-source personal assistant that operates across messaging platforms such as WhatsApp, Telegram, Slack, and Discord. Similar architectural patterns appear in platform SDKs from OpenAI~\citep{openai2025agentssdk}, Google~\citep{google2025adk}, and Anthropic~\citep{anthropic2024agents}, suggesting these challenges are broadly applicable. Unlike RAG systems, where interventions primarily affect content selection, agentic pipelines introduce multiple downstream decision points, including tool selection, action execution, memory updates, each of which can serve as an entry point for commercial influence.
 
\textbf{Intervention surfaces.} We highlight four stages in typical agentic pipelines (See Table~\ref{tab:agent_pipeline} that uses an OpenClaw-like architecture as an illustration):
\begin{enumerate}[nosep]
    \item \emph{Context assembly.} Each agent turn reconstructs the context window from sources such as bootstrap files, skill indices, tool schemas, and memory retrieval. \textbf{Challenge:} any party that influences these sources, by injecting brand associations into system prompts, skills, or memory, can shape all downstream responses. Because this assembly is not visible to the user, the resulting influence is effectively unobservable.
    \item \emph{Tool selection.} The agent selects which external API or service to invoke (e.g., Expedia vs.\ Booking.com) via function calling or the Model Context Protocol (MCP). \textbf{Challenge:} tool selection determines the entire downstream consideration set. If one provider is selected, alternatives are excluded entirely, i.e., a form of \emph{total} foreclosure that goes beyond the partial foreclosure documented in search self-preferencing~\citep{de2014integration,motta2023self}, where competitors can still appear (even if demoted).
    \item \emph{Memory persistence.} Agents maintain state across sessions through persistent memory. \textbf{Challenge:} commercial signals encoded in persistent memory affect all future interactions with no per-response cost or natural expiration, constituting Tier~4 influence by construction.
    \item \emph{Proactive scheduling.} Agentic frameworks support periodic and scheduled actions independent of user prompts. \textbf{Challenge:} these mechanisms enable agent-initiated product suggestions, introducing a Tier~3 surface with no analogue in passive LLM interfaces.
\end{enumerate}

\begin{table}[t]
\centering
\caption{Advertising entry points in an agentic pipeline (illustrated via OpenClaw-like architecture), mapped to influence tiers and the probabilistic framework.}
\label{tab:agent_pipeline}
\small
\begin{tabular}{@{}p{3cm}p{4.2cm}p{1.0cm}p{3.8cm}@{}}
\toprule
\textbf{Pipeline Stage} & \textbf{Advertising Opportunity} & \textbf{Tier} & \textbf{Shifted Variable} \\
\midrule
Context assembly (bootstrap files, skill indices, memory retrieval) & Brand associations in system prompts, persona files, or skill descriptions & 2, 4 & $P(X \mid u, c)$, $P(u' \mid u)$ \\
\addlinespace
Tool selection (MCP / function calling) & Which API the agent calls (e.g., Expedia vs.\ Booking.com) & 3 & $P(t \mid c, \text{Tools}, \mathbf{s})$ \\
\addlinespace
Tool execution & Search parameters, filters, result ordering within the selected tool & 2 & $P(X_t \mid c, t)$ \\
\addlinespace
Response generation & Which attributes to emphasize in the final recommendation & 1, 2 & $P(x \mid X_t, u, c)$ \\
\addlinespace
Memory write-back + heartbeat & Brand associations persisted via memory flush; proactive suggestions via cron/heartbeat & 3, 4 & $P(u' \mid u)$, $P(c' \mid c)$ \\
\bottomrule
\end{tabular}
\end{table}

\paragraph{The cascade challenge.} The critical structural challenge in agentic pipelines is the \textbf{cascade effect}: winning at an upstream stage (tool selection) can dominate all downstream outcomes. Formally:
\begin{equation}
P(x \mid u, c) = \sum_{t} \sum_{X_t} P(x \mid X_t, u, c) \cdot P(X_t \mid c, t) \cdot P(t \mid c, \text{Tools}, \mathbf{s}),
\end{equation}
where $t \in \text{Tools}$ denotes the selected tool and $X_t$ the information or options the agent provides. Interventions on $P(t \mid c,\text{Tools},\mathbf{s})$ determine which sources are accessible, constraining all downstream generation. This effect compounds in multi-stage pipelines: the output of an upstream tool becomes the input to downstream tools, and influence propagates recursively. Early interventions do not merely constrain a single decision but shape the entire trajectory of the interaction. Emerging empirical work supports these concerns: \citet{allouah2026your} audit AI shopping agents and find choice homogeneity, strong position biases varying across model providers, and susceptibility to seller manipulation of product descriptions; \citet{stockl2025ai} find that AI agents favor structured data over visual or emotional appeals, a fundamentally different engagement pattern from human users.
 
\textbf{Open problems.} The agentic setting raises challenges that remain largely unaddressed. How can we design mechanism frameworks for cascading, multi-stage pipelines? How can tool-selection foreclosure be mitigated without degrading agent functionality? How can persistent memory be audited when the memory stack is opaque to users? And how can influence be attributed when an upstream decision determines an entire downstream session? These challenges reflect the difficulty of governing influence in systems where interventions propagate across stages and persist over time.

\subsection{Towards Trustworthy Intervention in Generative Systems} 

Two challenges deserve emphasis even within the framework above. First, models that reject paid advertising %
eliminate platform-sanctioned ad interventions but do not eliminate commercial influence. Channels that operate without platform cooperation remain: third parties can still steer outputs through generative engine optimization~\citep{aggarwal2024geo}, adversarial content manipulation~\citep{nestaas2024adversarial}, or retrieval-time injection. Even systems with strong ad-free commitments therefore face a trustworthy-intervention problem at the input boundary: how to detect, attribute, and constrain commercially motivated influence entering through the broader generative pipeline. Second, the agentic setting introduces a reverse-inference channel: \citet{chen2026ads} show that LLMs can infer private user attributes, including political affiliation, employment status, and education level, from passive ad exposure patterns at substantially lower cost than human profiling. This compounds forward influence by making subsequent steering more targeted, and creates a measurement risk that grows as agentic personalization deepens.

\section{Discussion}
\label{sec:discussion}
Generative AI advertising extends far beyond the ``Sponsored'' cards appearing below ChatGPT responses today. Our taxonomy reveals that deployed systems and most academic mechanism design concentrate at the shallowest tier, i.e., explicit product placement, while the subtler forms (information framing, behavioral redirection, preference formation) are precisely those for which we have the fewest tools to detect, measure, or govern. Much of the evidence on Tiers~2--4 remains closer to feasibility demonstrations than mature deployments, and the associated business models and governance mechanisms remain nascent in a context whose dynamic personalization, expectation of objectivity, and expanding multi-modal and agentic surfaces make it qualitatively different from native advertising in search or social media.

\textbf{Our goal is not to propose a complete solution, but to provide a unifying framework for reasoning about commercial interventions in generative systems.} By organizing the design space into tiers and linking them to intervention points in the generative process, we make explicit where influence enters, how it propagates, and why it becomes harder to detect and govern at higher tiers. This framing shifts the research problem from optimizing placement to enabling \textbf{trustworthy intervention}, i.e., interventions that are observable, attributable, and aligned with user welfare.

Our taxonomy is primarily organized around text-based outputs; as multi-modal and agentic systems mature, the tier boundaries may shift or blur, and back-end agentic routing does not map cleanly onto any single tier. 
The framework provides organizing vocabulary but not formal identifiability results, and does not address domain-specific settings (medical, legal, financial) or multilingual contexts where detection difficulty may vary materially. We view this as a starting point for a research program, not a definitive mapping of the space.
 
As economic pressures push platforms toward deeper monetization and AI systems expand into multiple modalities and agents, the surface area for commercial influence may outpace our ability to study it. Developing methods for detection, attribution, and control at the latent and cascading layers is essential for sustaining user trust. And the work needs to start now, not after deployment.

\bibliographystyle{plainnat}
\bibliography{references}

\end{document}